\documentclass[twocolumn,tighten]{aastex63}
\usepackage{amssymb, amsmath,framed}
\usepackage{gensymb}
\usepackage{savesym}
\savesymbol{tablenum}
\usepackage{siunitx}
\restoresymbol{SIX}{tablenum}

\usepackage{bm}
\expandafter\ifx\csname package@font\endcsname\relax\else
 \expandafter\expandafter
 \expandafter\usepackage
 \expandafter\expandafter
 \expandafter{\csname package@font\endcsname}

\fi
\def\bq{\begin{equation}}
\def\eq{\end{equation}}
\def\bqy{\begin{eqnarray}}
\def\eqy{\end{eqnarray}}

\def\p{\partial}

\def\p{\partial}

\def\R{\mathbb{R}}

\begin{document}
\title{\large{Project Lyra: A Mission to 1I/'Oumuamua without Solar Oberth Manoeuvre}}

\correspondingauthor{Adam Hibberd}
\email{adam.hibberd@i4is.org}

\author{Adam Hibberd}
\affiliation{Initiative for Interstellar Studies (i4is) 27/29 South Lambeth Road London, SW8 1SZ United Kingdom}

\author{Andreas M. Hein}
\affiliation{Initiative for Interstellar Studies (i4is) 27/29 South Lambeth Road London, SW8 1SZ United Kingdom}

\author{T. Marshall Eubanks}
\affiliation{Space Initiatives Inc, Newport, VA 24128, USA}

\author{Robert G. Kennedy III}
\affiliation{Initiative for Interstellar Studies (i4is) 27/29 South Lambeth Road London, SW8 1SZ United Kingdom}

\begin{abstract}
 To settle the question of the nature of the interstellar object 1I/'Oumuamua requires in-situ observations via a spacecraft, as the object is already out of range of existing telescopes. Most previous proposals for reaching 1I/'Oumuamua using near-term technologies are based on the Solar Oberth Manoeuvre (SOM), as trajectories without the SOM are generally significantly inferior in terms of lower mission duration and higher total velocity requirement. While the SOM allows huge velocity gains, it is also technically challenging and thereby increases programmatic and mission-related risks. In this paper, we identify  an alternative route to the interstellar object 1I/'Oumuamua, based on a launch in 2028, which does not require a SOM but has a similar performance as missions with a SOM. It instead employs a Jupiter Oberth Manoeuvre (JOM) with a total time of flight of around 26 years or so. The efficacy of this trajectory is a result of it significantly reducing the $\Delta$V to Jupiter by exploiting the VEEGA sequence. The total $\Delta$V of the trajectory is 15.8 $\si{km.s^{-1}}$ and the corresponding payload mass is 115 $\si{kg}$ for a SLS Block 1B or 241 $\si{kg}$ for a Block 2. A further advantage of the JOM is that the arrival speed relative to 1I/'Oumuamua is approximately 18 $\si{km.s^{-1}}$, much lower than the equivalent for the SOM of around 30 $\si{km.s^{-1}}$.\\
 
\end{abstract}

\section{Introduction} \label{SecIntro}
The celestial body designated 1I/'Oumuamua is the first object to be discovered and identified as an interstellar object, stimulating much interest, debate and speculation from the scientific community \citep{bannister2019natural}. Theories to explain the nature of 1I/'Oumuamua have included a fractal dust aggregate \citep{Flekky2019}, a hydrogen iceberg \citep{Seligman2020}, a nitrogen iceberg \citep{Jackson2021,Desch2021}, an alien solar sail \citep{Bialy2018}, fragments of a tidally disrupted planet \citep{Raymond2018} and so on. All explanations have one feature in common - they are extraordinary. Given the enigmatic nature of 1I and to settle this matter, additional data would be vital. However, due to the large distance 1I has already traveled away, it is no longer possible to obtain data via telescopes. It has been argued that therefore, we should rather wait for the next similar object to come within range of telescopes.

However, the discovery of the second interstellar object 2I/Borisov in 2019 \citep{2019ApJ...886L..29J} confirmed that there is at least one interstellar object which resembles minor bodies found in the Solar System. This makes 1I even more of an oddity and it is unclear what the likelihood of encountering a similar object again is, although various estimates have been given in the literature \citep{portegies2018origin,bannister2019natural,Desch2021}. Hence, waiting for the next object similar to 1I to enter our Solar System might be fruitless. An alternative strategy would be to send a spacecraft to 1I. Despite initial statements against the feasibility of this option \citep{Bialy2018}, a series of publications has argued that such a mission would be feasible with near-term technologies \citep{Hein2019b,Seligman2018,HIBBERD2021594,HIBBERD2021,HIBBERD2020}.

As a contribution to this effort, previous research has been conducted into spacecraft missions to 1I/'Oumuamua \citep{Hein2019b,Seligman2018,HIBBERD2021594,HIBBERD2021,HIBBERD2020}, and the possible scientific return from such a venture make this an unmissable opportunity \citep{HEIN2022402,eubanks2020exobodies}. Given the limitations of present or near-term technology, these studies have generally identified a Solar Oberth Manoeuvre (SOM) as a requirement for viable spacecraft missions. As an exception, \cite{HIBBERD2020} have identified several trajectories without a SOM. However, these trajectories seem to be generally inferior in terms of longer mission duration and higher required $\Delta$V, compared to missions with a SOM. 

While the advantages of the SOM have been demonstrated, the recent Interstellar Probe endeavour has highlighted the multiple drawbacks of such a manoeuvre \citep{InterstellarProbe} and in its place has proposed either 1) a passive gravitational assist of Jupiter, or 2) a powered GA of Jupiter (a Jupiter Oberth), the SOM being the least preferred.

While the arguments against the SOM can be debated, investigating the prospect of a mission to 1I without the SOM but comparable performance seems to merit consideration, as it would certainly alleviate some of the technological challenges of the SOM such as the development of a scaled up Parker Solar Probe heat shield and a precise thrust orientation during the Manoeuvre. Without the SOM, the mission would much more resemble existing interplanetary missions. 

In this context, here we present an alternative solution, a mission to 'Oumuamua which has relatively low $\Delta$V, yet does not conduct a SOM to accelerate the craft. The mission is abbreviated as V-E-DSM-E-J , adopting the convention that the home planet (E) and target body (1I) are implicitly assumed. Usually in the context of a mission to Jupiter, this is alternatively referred to as VEEGA.

\section{Method}

As in previous Project Lyra studies, The 'Optimum Interplanetary Trajectory Software' (OITS) \citep{OITS_info} was employed to research missions to 'Oumuamua. It adopts the following assumptions:

\begin{enumerate}
    \item a patched conic for every planet encountered
    \item the thrust is impulsive resulting in discrete applications of $\Delta$V along the trajectory
    \item the $\Delta$V is applied at the point of periapsis w.r.t. every planet encountered
    \item this $\Delta$V is applied tangential to the motion of the spacecraft at every planet encountered
    \end{enumerate}

As an alternative instead of a planet, an 'Intermediate Point' may be chosen, a point whose heliocentric longitude and latitude are optimised by OITS but whose sun distance is user-specified. This can be selected to model a Deep Space Manoeuvre (DSM). 

Two Non-Linear Programming (NLP) optimizers were used for this study, firstly NOMAD \citep{LeDigabel2011} for a general solution and to fine-tune the results MIDACO \citep{Schlueter_et_al_2009,Schlueter_Gerdts_2010,Schlueter_et_al_2013}.

\section{Results}

\begin{table*}[]
\centering
\caption{Mission V-E-DSM-E-J (VEEGA) with 22 Years Flight Duration Constraint}
\label{table:1}
\begin{tabular}{|c|c|c|c|c|c|c|c|}
\hline
Num & Planet       & Time        & Arrival speed & Departure speed & $\Delta$V  & Cumulative $\Delta$V & Periapsis \\
    &              &             & \si{km.s^{-1}}          & \si{km.s^{-1}}            & \si{km.s^{-1}}    & \si{km.s^{-1}}              & \si{km}        \\ \hline
1   & Earth        & 2028 FEB 04 & 0.0000             & 4.1973          & 4.1973  & 4.1973            & N/A       \\ \hline
2   & Venus        & 2028 JUN 23 & 6.7660         & 5.0616          & 0.8536  & 5.0509            & 200.0       \\ \hline
3   & Earth        & 2029 FEB 27 & 7.5338        & 7.5338          & 0.0000       & 5.0510             & 2900.2    \\ \hline
4   & DSM @ 2.2 au & 2030 JAN 24 & 15.8493       & 15.4643         & 0.4184  & 5.4693            & N/A       \\ \hline
5   & Earth        & 2031 FEB 09 & 9.8911        & 12.9506         & 2.1978  & 7.6671            & 200.0       \\ \hline
6   & Jupiter      & 2032 APR 17 & 16.8277       & 41.1420          & 10.5122 & 18.1793           & 200.0       \\ \hline
7   & 'Oumuamua    & 2050 JAN 29 & 22.4983       & 22.4983         & 0.0000       & 18.1793           & N/A       \\ \hline
\end{tabular}
\end{table*}
\begin{table*}[]
\centering
\caption{Mission V-E-DSM-E-J (VEEGA) with 26 Years Flight Duration Constraint}
\label{table:2}
\begin{tabular}{|c|c|c|c|c|c|c|c|}
\hline
Num & Planet       & Time        & Arrival speed & Departure speed & $\Delta$V  & Cumulative $\Delta$V & Periapsis \\
    &              &             & \si{km.s^{-1}}          & \si{km.s^{-1}}            & \si{km.s^{-1}}    & \si{km.s^{-1}}              & \si{km}        \\ \hline
1 & Earth        & 2028 MAR 07 & 0.0000                       & 3.4550                      & 3.4550          & 3.4550                     & N/A            \\ \hline
2 & Venus        & 2028 JUL 06 & 5.6526                  & 5.6511                    & 0.0070             & 3.4557                     & 200.0            \\ \hline
3 & Earth        & 2029 MAR 13 & 8.0443                  & 6.2791                    & 0.9708        & 4.4265                   & 393.6          \\ \hline
4 & DSM @ 2.2 au & 2030 FEB 12 & 15.9254                 & 15.4781                   & 0.4518        & 4.8783                   & N/A            \\ \hline
5 & Earth        & 2031 FEB 15 & 9.0627                  & 13.2924                   & 3.0000        & 7.8783                   & 200.0            \\ \hline
6 & Jupiter      & 2032 MAR 22 & 18.5791                 & 37.0710                   & 7.9588        & 15.8371                  & 4464.5            \\ \hline
7 & 'Oumuamua     & 2054 MAR 01 & 18.1348                 & 18.1348                   & 0.0000             & 15.8371                  & N/A            \\ \hline
\end{tabular}
\end{table*}
\begin{table*}[]
\centering
\caption{Mass Breakdown for Jupiter Oberth, assuming a total Jupiter $\Delta$V of 8.0 \si{km.s^{-1}}}
\label{table:3}
\begin{tabular}{|c|c|c|c|c|}
\hline
        &          &         & Stage2   &          \\ \hline
        &          & STAR 75 & STAR 63F & STAR 48B \\ \hline
        & STAR 75  & 342 kg     & 406 kg     & 378 kg     \\
        &          & 16478 kg   & 13064 kg   & 10580 kg    \\ \hline
Stage 1 & STAR 63F & 169 kg   & 239 kg    & 241  kg    \\
        &          & 12827 kg   & 9419 kg    & 6966 kg    \\ \hline
        & STAR 48B & 26 kg     & 94 kg      & 115 kg     \\
        &          & 10228 kg  & 6818  kg   & 4383 kg    \\ \hline
\end{tabular}
\end{table*}

\begin{figure*}[h]
\centering
\includegraphics[scale=0.5]{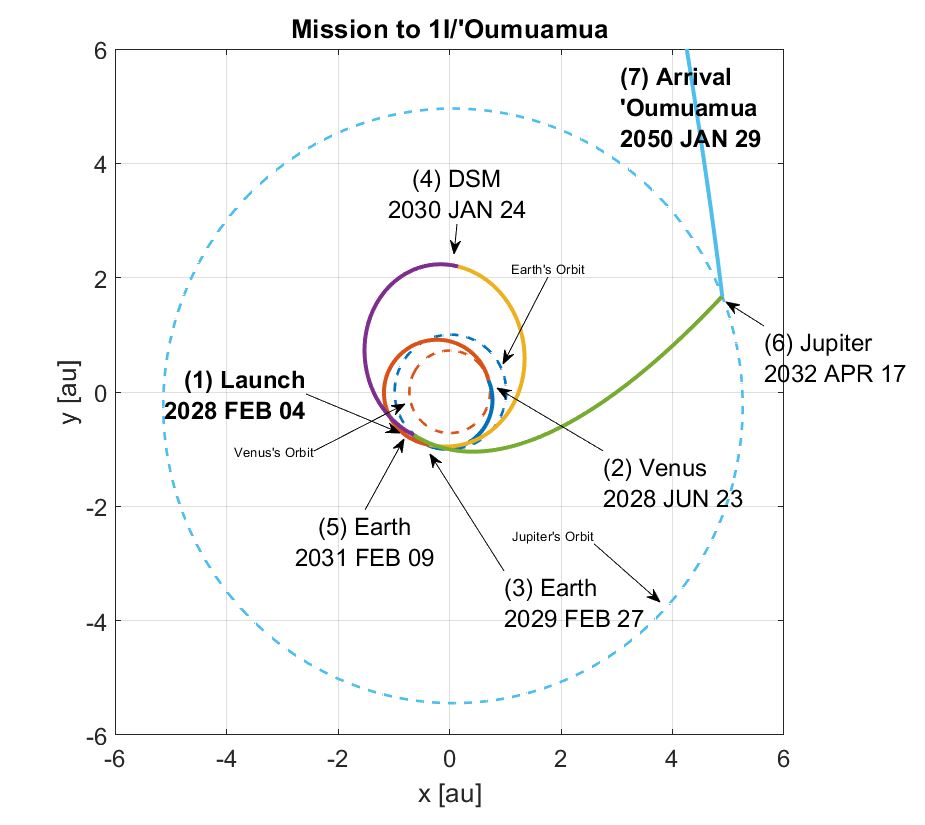}
\caption{Trajectory, V-E-DSM-E-J (VEEGA)}
\label{fig:Trajectory}
\end{figure*}

\begin{table*}[]

\centering
\caption{Launch Vehicle Data for $C_{3} = 12 \si{km^{2}. s^{-2}}$}
\label{table:4}
\begin{tabular}{|c|c|c|}
\hline
Launch Vehicle          & Payload Mass to $C_{3} = 12 \si{km^{2}. s^{-2}}$  (kg) & Remaining Mass for JOM (kg) \\ \hline
SLS Block 1B            & 33000                           & 6670                        \\ \hline
SLS Block 2            & 43000                           & 8690                        \\ \hline
Falcon Heavy Expendable & 11880                           & 2401                        \\ \hline
\end{tabular}
\end{table*}
The solution trajectory is shown in Figure \ref{fig:Trajectory} (coinciding with the 2028 VEEGA trajectory in \cite{doi:10.2514/2.3650}) and the numerical information is provided in Table \ref{table:1} . This scenario amounts to the same VEEGA sequence to Jupiter  adopted for instance by the Galileo spacecraft, where it was shown to be the globally optimal trajectory for the mission \citep{doi:10.2514/2.3800}. Furthermore \cite{Zubco2021} also exploit this sequence in their analysis of missions to Sedna. Note the mission duration constraint of 22 years (as also imposed in \cite{HIBBERD2020}) results in a $\Delta$V of 10.5 \si{km.s^{-1}} delivered at Jupiter (row 6). This is rather larger than is easily reached by two staged solid rocket motors (SRMs), so in order to reduce this $\Delta$V, the flight duration restriction may be relaxed.

\begin{figure*}[h]
\centering
\includegraphics[scale=0.5]{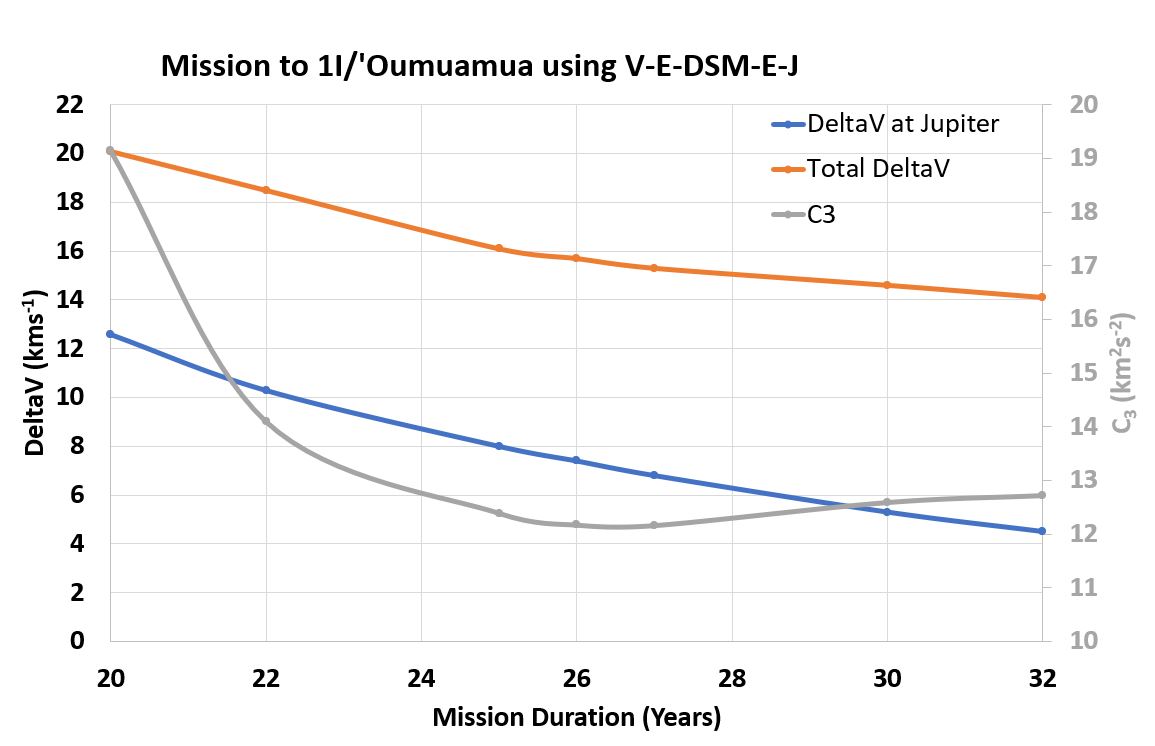}
\caption{ $\Delta$V vs Flight Duration, V-E-DSM-E-J (VEEGA)}
\label{fig:ToF}
\end{figure*}

\begin{figure}[h]
\hspace*{-1.4cm}
\includegraphics[scale=0.52]{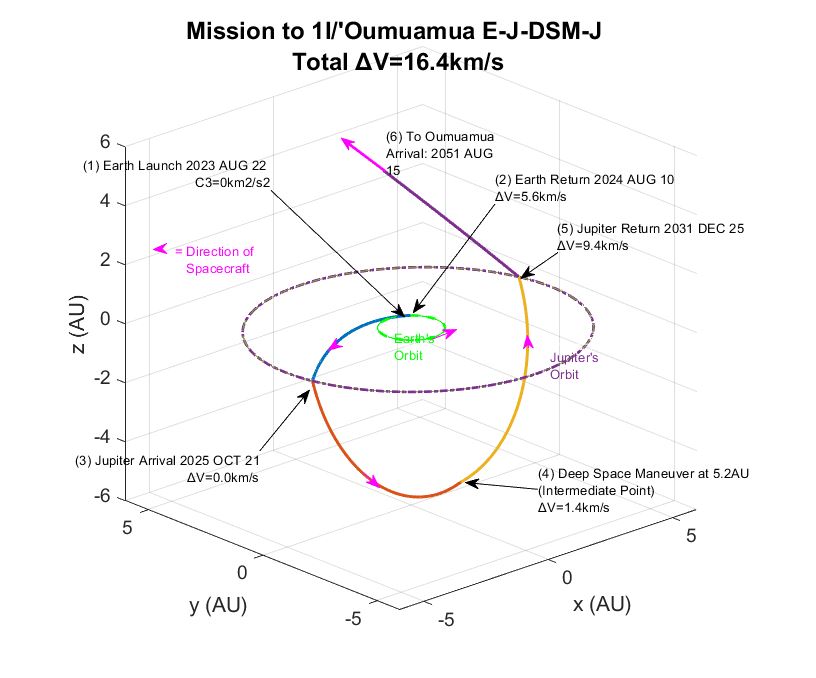}
\caption{Trajectory, E-J-DSM-J}
\label{fig:Trajold}
\end{figure}

Figure \ref{fig:ToF} is a graph revealing the dependency of overall $\Delta$V and the Jupiter $\Delta$V on mission duration constraint for the V-E-DSM-E-J, VEEGA scenario. On the secondary vertical axis we have the magnitude of $C_{3}$ (equivalent to the square of the hyperbolic excess speed at Earth departure - row 1 of Table \ref{table:1}).

From this Figure we adopt the 26 year mission duration as it approximately coincides with the minimum $C_{3}$ at launch. In order to ensure the in-flight $\Delta$V allows room for a significant payload to be delivered to 1I/'Oumuamua, an extra constraint was applied to the $\Delta$V at the second Earth gravitational assist. Thus refer to Table \ref{table:2} wherein we see a constraint of 3.0 \si{km.s^{-1}} is imposed at this encounter (row 5). An animation of this trajectory can be found at \cite{CiteTrajAnim}.

Addressing the Jupiter Oberth Manoeuvre (JOM- Row 6 of Table \ref{table:2}), we observe to deliver such a hefty kick in velocity, 8.0 \si{km.s^{-1}}, we need as a minimum 2 staged SRMs. We assume some combination of STAR 75, STAR 63F and STAR 48B is exploited, the situation is summarised in Table \ref{table:3}. All the possible permutations of these SRMs are accounted for and we find that the payload mass for any combination is positive.

Preceding the JOM, there is a total of 4.4 \si{km.s^{-1}} of $\Delta$V applied en route. Let us assume that all of this $\Delta$V is delivered by a throttleable Liquid Propellant Engine (assuming $MMH/N_2O_4$ fuel, $I_{sp}$=341 \si{s}, $p=0.1$). In addition we find that for the 26 year duration mission, there is a $C_{3}$ value at Earth of $3.455^2 = 11.94$ \si{km^{2} s^{-2}}. Table \ref{table:4} provides the payload mass capability of 3 launch vehicles to a $C_{3}$=12 \si{km^{2} s^{-2}} escape orbit, the Space Launch System (SLS) Block 1B, the SLS Block 2 and a Falcon Heavy Expendable. 

Referencing Tables \ref{table:3} and \ref{table:4}, we find that the Block 1B with two staged STAR 48B SRMs can deliver 115 \si{kg} to 'Oumuamua, whereas the Block 2 is capable of 241 kg, with a STAR 63F followed by a STAR 48B. The Falcon Heavy Expendable, however, is unable to achieve enough payload to deliver the necessary JOM $\Delta$V, even assuming 2 STAR 48B SRMs.

At this juncture as a side note, we reiterate a previously discovered trajectory in \cite{HIBBERD2020}. It is depicted in Figure \ref{fig:Trajold} and can be abbreviated as E-J-DSM-J. The DSM is placed at 5.2 $\si{au}$ from the sun and at a low (negative) heliocentric latitude. The relevance is that it is also an option which does not require a SOM, and in fact it exploits 2 JOMs in succession, separated by half a Jupiter cycle, i.e. approximately 6 years. The launch would have to be be in 2023, which is unfortunately too soon to organise a mission using this scenario.

\section{Discussion}
We have demonstrated that at least one trajectory for a mission to 1I exists without a SOM. A mission with such a trajectory has multiple advantages: 
\begin{itemize}
\item The development of a scaled up version of the Parker Solar Probe heat shield can be omitted, thereby decreasing development risk.
\item The mass of the heat shield is thereby removed from the mass budget, freeing up mass that can be dedicated to other spacecraft subsystems.
\item The mission risk of a precise steering of the thrust vector during the Oberth Manoeuvre is eliminated. The velocity vector at the end of the Oberth Manoeuvre is very sensitive to the thrust vector applied to the spacecraft during the Manoeuvre. Even small errors can lead to a large deviation of the spacecraft from its target trajectory. 
\item The mission much closer resembles the existing heritage of interplanetary missions and decreases its implementation risk.

\item The arrival velocity of the spacecraft w.r.t. 1I is lower (18  $\si{km.s^{-1}}$) than that for the SOM (30 $\si{km.s^{-1}}$).
\end{itemize}

There are also disadvantages with missions to 1I without an SOM:

\begin{itemize}
    \item Due to the large number of flybys, launch opportunities are much rarer than for missions with an SOM.
    \item The mission duration is longer than for missions with an SOM: about 25 years in the present case versus mission durations of $<$ 20 years. 
\end{itemize}


\section{Conclusion}
In this paper, we presented a mission and trajectory design to 1I/'Oumuamua without a Solar Oberth Manoeuvre but comparable performance. Our results indicate that with a combination of Venus, Earth, and Jupiter flybys, including a Deep Space Manoeuvre, a mission to 1I can be staged in 2028 with reaching 1I in 2054. The total $\Delta$V of the mission would be 15.8 \si{km.s^{-1}}. The absence of a Solar Oberth Manoeuvre significantly reduces the development risks of such a mission, mainly due to elimination of the heat shield. The freed up mass can be allocated to other spacecraft subsystems. Also, the specific mission risks of a Solar Oberth Manoeuvre, such as the precise burn and velocity vector orientation at the end of the Manoeuvre are alleviated. However, a clear disadvantage is that launch opportunities for such a mission are much rarer and therefore reduce the flexibility. Also, the complexity of the mission with its multiple flybys is increased. This may, nevertheless, not be a significant drawback, as current interplanetary missions face the same issues. Finally, our results rather strengthen the feasibility of a mission to 1I, as we demonstrated that such a mission may resemble existing interplanetary missions and not rely on significant new technology developments or unprecedented space Manoeuvres. 

\bibliographystyle{aasjournal}
\bibliography{library,library_Adam_Hibberd,Hein_ISO_Modified_by_Adam_Hibberd}

\end{document}